# Outlier-immune Data-driven Linear Power Flow Model Construction via Mixed-Integer Programming

Guoan Yan, *Student Member, IEEE*, Zhengshuo Li, *Senior Member, IEEE*

*Abstract*—The common approaches to construct a data-driven linear power flow (DD-LPF) model cannot completely eliminate the adverse impacts of outliers in a training dataset. In this letter, a novel outlier-immune DD-LPF model construction method via mixed-integer programming is presented for automatically and optimally identifying outliers to form a more accurate LPF model. Two acceleration solution strategies are further suggested to reduce the computational time. Case studies demonstrate the superior accuracy and comparable computational time of the proposed method when compared to three common approaches.

## I. INTRODUCTION

THE AC power flow constraint renders many power system optimization problems NP-hard. Constructing a relatively accurate linear power flow model to replace the AC power flow constraints is promising for circumventing this NP-hard issue and has received much attention [1], [3]-[8]. Among construction methods, data-driven approaches, which construct a data-driven linear power flow (DD-LPF) model via regression methods, have recently become highly regarded [1].

One challenge in constructing an accurate DD-LPF model is handling outliers, or *bad data*, in the training dataset, especially when the bad data do not notably differ from the other data. Among the common regression methods, the well-known least squares (LS) method is vulnerable to outliers [2], while the least absolute value (LAV) method cannot always guarantee a satisfying outcome [3]. Moreover, despite having been tested in some works, methods employing support vector regression (SVR) [4], the Huber penalty function [5], [6], or the use of a two-stage procedure to implement the largest normalized residual (LNR) bad data identification [7] cannot theoretically guarantee that the resultant fitting model is *completely* shielded against outliers. This can be roughly explained by the fact that these common methods cannot strictly insulate the fitting process from outliers that are mixed with other data and are difficult to distinguish and screen out *a priori*.

To construct an outlier-immune DD-LPF model, a novel construction method via mixed-integer programming (MIP) is proposed in this letter. Specifically, since the statistics of the ratio of bad data to the whole dataset are usually known by practitioners, one can introduce data sample-related binary variables to formulate an MIP fitting model that is subject to the known (or readily estimated) bad data ratio. This model can thus **automatically** and **optimally** identify the most suspectable data and eliminate their adverse impact on the LPF model constructed. Hence, we call this the *outlier-immune* DD-LPF construction. Despite the risk of excluding a small portion of normal data when using a conservative outlier ratio estimation, case studies demonstrate that the proposed method can always construct a more accurate DD-LPF model than the two-stage LNR, SVR or the LS-Huber penalty combined (LHC) approach against different outlier ratios. Moreover, given a relatively large training dataset that entails increasing binary variables, two acceleration solution strategies are suggested, which are confirmed by case studies to significantly reduce the computing time, thus enhancing the practicality of the proposed method.

## II. PRELIMINARIES

To better explain our method, a typical DD-LPF model is first introduced, followed by two common outlier-handling approaches. Generally, a DD-LPF model can be formulated as $y = \hat{\mathbf{A}}(\hat{\boldsymbol{\omega}})x + \hat{\boldsymbol{b}}(\hat{\boldsymbol{\omega}})$, i.e., a linear relationship between the vector of the independent variable $x$ (e.g., nodal voltages) and the vector of dependent variables $y$ (e.g., nodal power), where the matrix $\hat{\mathbf{A}}(\hat{\boldsymbol{\omega}})$ and the vector $\hat{\boldsymbol{b}}(\hat{\boldsymbol{\omega}})$ are linearly determined by the parameter vector $\hat{\boldsymbol{\omega}}$. By expanding $\hat{\boldsymbol{\omega}}$ into $\boldsymbol{\omega} = (1 \quad \hat{\boldsymbol{\omega}})^T$, $y = \hat{\mathbf{A}}(\hat{\boldsymbol{\omega}})x + \hat{\boldsymbol{b}}(\hat{\boldsymbol{\omega}})$ can be further reformulated as $y = \mathbf{A}(x)\boldsymbol{\omega}$, which is to be fitted, e.g., with an LS method, over the historical training dataset $\{(x_1, y_1), (x_2, y_2), \cdots (x_m, y_m)\}$:

$$\boldsymbol{\omega} = \arg\min_{\boldsymbol{\omega}} \sum_{i=1}^{m} \|y_i - \mathbf{A}(x_i)\boldsymbol{\omega}\|_2^2. \qquad (1)$$

A common method for addressing outliers is LHC [5], which replaces the L2-norm residuals $r_i(\boldsymbol{\omega}) = \|y_i - \mathbf{A}(x_i)\boldsymbol{\omega}\|_2$ in (1) with $\phi_{hub}(r_i(\boldsymbol{\omega}))$ below:

$$\phi_{hub}(r_i(\boldsymbol{\omega})) = \begin{cases} r_i(\boldsymbol{\omega})^2 & r_i(\boldsymbol{\omega}) \leq \delta \\ \delta(2r_i(\boldsymbol{\omega}) - \delta) & r_i(\boldsymbol{\omega}) > \delta \end{cases}. \qquad (2)$$

Since the effect of residuals larger than the Huber threshold $\delta$ changes from quadratic to linear, the impact of outliers on the fitting result is reduced by the LHC approach.

Another typical approach to handling outliers is the two-stage LNR procedure, as described in [7]. Briefly, this procedure is conducted as follows:

**Stage 1**: (1) is solved, and the residual covariance matrix is calculated to obtain the normalized residuals.

**Stage 2**: If the *i*-th normalized residual is larger than the

This work is supported by National Key R&D Program of China under grant 2022YFB2402900. G. Yan, and Z. Li are with Shandong University, Jinan 250061, China. (Corresponding author: Zhengshuo Li.)

identification threshold, then the associated data are suspected to be outliers and removed before returning to Stage 1; otherwise, the process is terminated.

This two-stage LNR process, as [9] notes, has a significant shortcoming: "Successive elimination of largest normalized residual measurements may result in wrong suppressions of good measurements". In addition, this procedure is heuristic, and the identification threshold is artificially determined; as a result, this method cannot theoretically ensure that the impact of outliers on the fitting result is completely eliminated.

### III. THE PROPOSED METHOD

#### A. Original Formulation

The above analysis reveals that the adverse impact of outliers cannot be completely eliminated by common methods. To fix this issue, a straightforward idea is to introduce binary variables associated with the data sample as an indicator of whether the data are outliers to be excluded. In other words, we no longer need to empirically identify possible bad data in a sequential or heuristic way (e.g., in the two-stage procedure). Instead, we treat all the data as equal candidates for possible outliers and employ the following MIP formulation to find the **optimal identification** subject to the predefined ratio of outliers that can be available via field operation statistics.

$$\min_{\boldsymbol{\omega},z} \quad \sum_{i=1}^{m} \left\| \boldsymbol{y}_i - \mathbf{A}(\boldsymbol{x}_i)\boldsymbol{\omega} \right\|_2^2 (1-z_i) \quad (3)$$
$$\text{s.t.} \quad z \in \{0,1\}^m, \left\| z \right\|_1 \leq p \cdot m,$$

where $z_i = 1$ indicates that the associated sample $\boldsymbol{x}_i, \boldsymbol{y}_i$ is detected as an outlier, $z_i = 0$ indicates that the associated sample can be used to train the fitting model, $m$ is the volume of training data, and $p$ is the predefined outlier ratio.

**Discussion 1**: Given $p$, Model (3) can **automatically** identify the most suspicious data and completely eliminate their adverse impact due to the term $(1-z_i)$. Moreover, the solution $z^*$ is **optimal** in the sense that the residuals of the fitting model must be the minimum over the training set. Granted, given a conservative estimation of the outlier ratio, i.e., $p$ is larger than the actual ratio of the outliers, a small portion of the normal data is excluded from fitting. However, as the training dataset volume is usually sufficient for LPF model fitting, the detriment of this conservative act is usually negligible.

**Discussion 2**: In practice, typically only some of the components of the vectors $\boldsymbol{x}_i, \boldsymbol{y}_i$ can be outliers. In this sense, a natural approach is to associate every component of the vectors with a binary variable. However, this approach introduces too many binary variables. Hence, given that $m$ is usually large enough for fitting, model (3) directly associates $\boldsymbol{x}_i, \boldsymbol{y}_i$ with $z_i$ to reduce the number of binary variables and decrease the computational burdens at the slight cost of discarding the "good" portions of the outlier vectors.

Model (3) has a cubic term, so the reformulation technique in [10] is introduced to obtain the equivalent mixed-integer quadratic programming (MIQP) model below:

$$\min_{\boldsymbol{\omega},z} \quad \sum_{i=1}^{m}\sum_{j=1}^{n} u_{ij}^2,$$
$$\text{s.t.} \quad -y_{ij} + \boldsymbol{a}_{ij}\boldsymbol{\omega} \leq u_{ij} + Mz_i, \; y_{ij} - \boldsymbol{a}_{ij}\boldsymbol{\omega} \leq u_{ij} + Mz_i, \quad (4)$$
$$\left\| z \right\|_1 \leq p \cdot m, z \in \{0,1\}^m$$

where $\boldsymbol{a}_{ij}$ is the $j$-th row of $\mathbf{A}(\boldsymbol{x}_i)$, $M$ is a large constant (e.g., $10^6$) and $u_{ij}$ is the auxiliary variable.

#### B. Iterative Conic Relaxation Acceleration Strategy (S1)

Although model (4) can be directly solved by off-the-shelf solvers, the solution time can be prohibitively long when a large training dataset is used because the same number of binary variables is also used. To address this issue, the iterative conic relaxation acceleration solution proposed in [11], called **Strategy S1** in this letter, can be employed. The idea of S1, as delineated in Fig. 1, is to split the above MIQP problem into inner and outer continuous QP subproblems via conic relaxation reformulation. By iteratively solving these problems, one can identify a better starting point to rapidly solve the MIQP (4) in the final step.

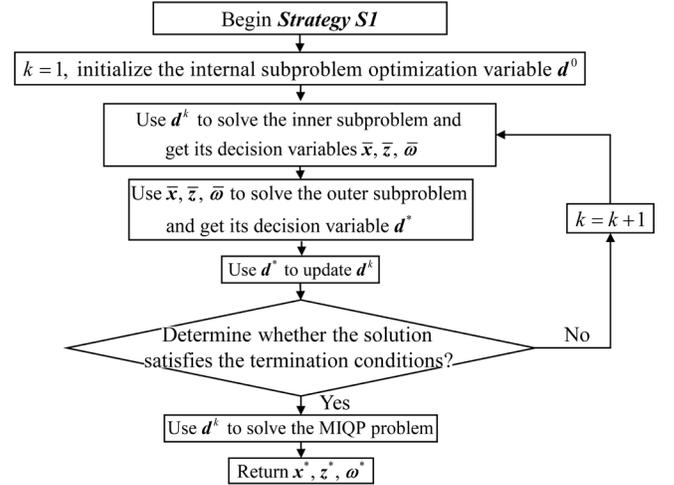

Fig. 1 Flowchart of Strategy S1.

**Discussion 3**: Ref. [11] has shown that S1 can often solve (4) faster than the off-the-shelf solvers, although a mixed-integer program should still be employed in the final step. We attribute this to the fact that a better starting point for searching for optimal binary variables is obtained through conic relaxation steps. Nevertheless, this strategy still requires inner and outer subproblems to be iteratively solved. Moreover, solving the mixed-integer program in the final step could be slower than expected if the starting point given by the previous steps is not good enough. This motivates us to propose an alternative strategy below.

#### C. Mixed-integer Linear Programming-based Acceleration Strategy (S2)

By replacing the L2-norm in (3) with an L1-norm and using big-M techniques, one can obtain the MILP model below:

$$\min_{\boldsymbol{\omega},z,s} \quad \sum_{i=1}^{m} \left\| \boldsymbol{y}_i - \mathbf{A}(\boldsymbol{x}_i)\boldsymbol{\omega} + \boldsymbol{s}_i \right\|_1 \quad (5)$$
$$\text{s.t.} \quad \left\| z \right\|_1 \leq p \cdot m, \; -Mz_i \leq s_i \leq Mz_i, z \in \{0,1\}^m.$$

The slack variables $s_i$ ensure that at most $p \cdot m$ outliers are being immunized and that $s_{i*}$ represents each element in $s_i$.

**Discussion** 4: Given the same number of binary variables, this MILP model is easier to solve than Model (4). Incidentally, since the L1-norm is more "robust" against noise than the L2-norm, Model (5) can often be expected to yield a slightly more accurate result over the training set (but not necessarily for the test set). This is shown in the case study.

## IV. CASE STUDIES

Extensive case studies were conducted to evaluate the performance of the proposed method on the IEEE three-phase unbalanced 123-node system and the IEEE balanced 300-bus system. The codes were run on a PC with an Intel i7-13700F 2.10 GHz CPU and 64 GB of RAM. The computational performances of SVM [4], LHC [5], two-stage LNR [7], and three versions of the proposed method, including directly solving model (4) and the S1 and S2 strategies, are compared. Two metrics of accuracy, maximum and average relative errors of the LPF model constructed, were calculated following [4]. The training and test sets were independently generated for each case study, and other simulation details were established according to [5]; these details are omitted to save space.

The first test was conducted on a 123-node system in response to Discussion 1; this test was conduct to examine the accuracy of the LPF model constructed by the proposed method when the *estimated* outlier ratio $p$ in (3) deviates from the *actual* outlier ratio $p_0$ (set to 8% according to [9]). Table I presents the results under the scenarios where $p$ is below, equal to, or above $p_0$. The results suggest that, when compared to the case $p=4\% < p_0$, conservatively estimating the outlier ratio (e.g., $p=12\%$) does **not** notably compromise model accuracy despite the risk of excluding a small portion of normal data. Since this condition is rather easy to meet in practice, this indicates that our method will **not** impose additional hard requirements upon the practitioner.

TABLE I
MAXIMUM/AVERAGE RELATIVE ERRORS OF THE LPF MODEL IN THE UNBALANCED 123-NODE SYSTEM WITH DIFFERENT $p$ (UNIT: %)

| $p_0=8\%$ | $p=4\% < p_0$ | $p=8\% = p_0$ | $p=12\% > p_0$ | $p=16\% > p_0$ |
|---|---|---|---|---|
| Training | **0.93**/0.51 | **0.55**/0.30 | **0.56**/0.30 | **0.57**/0.30 |
| Test | **1.23**/0.62 | **0.76**/0.37 | **0.78**/0.37 | **0.81**/0.38 |

Because the LPF model accuracy reduction is negligible when $p$ goes beyond $p_0$ and to save space, we present the results only when $p$ is equal to $p_0$ below. All three DD-LPF construction methods and two strategies, S1 and S2, are compared in two systems with $p_0=6\%$ and 10%, respectively. The results are summarized in Tables II-IV.

In terms of accuracy, the relative errors of the LPF models by the proposed method are notably smaller than those by the other methods, especially for larger $p_0$ values. In terms of computing time, the LNR requires multiple iterations between two stages. Directly solving Model (4) also takes a long time, but S2 requires the same order of time as the LHC and SVR approaches. Hence, after using Strategy S2 or S1, the proposed method yields a more accurate LPF model in a comparable computing time.

TABLE II
MAXIMUM/AVERAGE RELATIVE ERRORS OF THE FIVE LPF MODELS BY THE FIVE METHODS FOR UNBALANCED 123-NODE SYSTEM (UNIT: %)

| Outlier proportions $p_0$ | | Two-stage LNR | LHC | SVR | Directly solving (4) | S1 | S2 |
|---|---|---|---|---|---|---|---|
| Training | 6% | **0.88**/0.52 | **1.12**/0.57 | **0.83**/0.45 | **0.66**/0.34 | **0.55**/0.32 | **0.54**/0.30 |
| | 10% | **0.97**/0.56 | **1.75**/0.88 | **1.15**/0.58 | **0.68**/0.34 | **0.57**/0.32 | **0.56**/0.30 |
| Test | 6% | **1.45**/0.90 | **1.72**/0.85 | **1.32**/0.67 | **0.87**/0.52 | **0.71**/0.41 | **0.74**/0.37 |
| | 10% | **1.61**/0.97 | **3.05**/1.44 | **1.77**/0.86 | **0.90**/0.53 | **0.75**/0.43 | **0.77**/0.38 |

TABLE III
MAXIMUM/AVERAGE RELATIVE ERRORS OF THE FIVE LPF MODELS BY THE FIVE METHODS FOR BALANCED 300-BUS SYSTEM (UNIT: %)

| Outlier proportions $p_0$ | | Two-stage LNR | LHC | SVR | Directly solving (4) | S1 | S2 |
|---|---|---|---|---|---|---|---|
| Training | 6% | **1.51**/0.84 | **3.39**/1.36 | **1.43**/0.77 | **1.16**/0.60 | **1.12**/0.61 | **1.12**/0.62 |
| | 10% | **1.83**/0.97 | **4.35**/1.93 | **2.26**/1.14 | **1.23**/0.65 | **1.18**/0.64 | **1.16**/0.62 |
| Test | 6% | **3.78**/1.99 | **5.83**/2.71 | **1.88**/0.98 | **1.43**/0.75 | **1.38**/0.75 | **1.36**/0.73 |
| | 10% | **3.99**/2.12 | **7.11**/3.32 | **2.67**/1.31 | **1.48**/0.82 | **1.40**/0.81 | **1.42**/0.75 |

TABLE IV
THE COMPUTATIONAL TIME OF THE FIVE METHODS IN THE IEEE THREE-PHASE UNBALANCED 123-NODE SYSTEM (UNITS: SECOND)

| Outlier proportions $p_0$ | Two-stage LNR | LHC | SVR | Directly solving (4) | S1 | S2 |
|---|---|---|---|---|---|---|
| 6% | 39.29 | 5.64 | 3.31 | 715.91 | 11.39 | 3.21 |
| 10% | 52.42 | 4.92 | 3.74 | 889.78 | 27.52 | 7.29 |

## V. CONCLUSION

A novel outlier-immune DD-LPF construction method is proposed to optimally eliminate the adverse impact of outliers on the LPF model constructed. Two acceleration strategies are proposed and validated through simulations that they achieve superior accuracy and comparable computational time to the common methods. Future work will focus on testing parallel computing techniques to further reduce the computational time of S1 and S2.